\documentclass[prl, superscriptaddress, floatfix, twocolumn]{revtex4}

\usepackage{oldgerm}
\usepackage{latexsym}
\usepackage{amsmath}
\usepackage{amssymb}
\usepackage{amsgen}
\usepackage{amscd} 
\usepackage{color}
\usepackage{epsfig}
\usepackage{amsfonts}

\newcommand{\opr}[1]{\operatorname{#1}}

\def\one{{\mathchoice{\rm 1\mskip-4mu l}{\rm 1\mskip-4mu l}{\rm 1\mskip-4.5mu l}{\rm
1\mskip-5mu l}}}

\begin{document}

\title{Comment on ``Operator Quantum Error Correction"}

\author{Gerald Gilbert, Michael Hamrick,  F. Javier Thayer and Yaakov S. Weinstein\\
%\affiliation{
{\it Quantum Information Science Group}\\
{\rm MITRE}\\
{\it 260 Industrial Way West, Eatontown, NJ 07724 USA}}
%}

\begin{abstract}
The attempt to equate operator quantum error correction (quant-ph/0504189v1) with the quantum computer condition
(quant-ph/0507141) in version two of quant-ph/0504189 is shown to be invalid.

\end{abstract}

\maketitle
%\section{Introduction}

After the appearance of our paper~\cite{GHT2005}, the authors of~\cite{KLPLv1} released a 
revised version ({\em {cf}} ~\cite{KLPLv2}), in which they make the following incorrect assertions 
about our paper:
\vspace{.4cm}
\paragraph{\bf {Assertion 1:}}Equation (8) in our paper~\cite{GHT2005} is
``the fundamental formula in the formulation of the `Quantum Computer
Condition'~" ({\em {cf}}~ the second sentence in the third paragraph of~\cite{KLPLv2});
\vspace{.4cm}
\paragraph{\bf {Assertion 2:}}
Equation (8) in our paper~\cite{GHT2005} is ``captured as a special case of the UNS
framework" ({\em {cf}}~ again the second sentence in the third paragraph of~\cite{KLPLv2}).
\vspace{.4cm}

In this Comment we demonstrate that the above assertions are erroneous.
\vspace{.425cm}
\paragraph{\bf Response to 1:} Equation (8) in our paper ~\cite{GHT2005} is
\begin{equation}
\mathcal{M}_\mathrm{dec}(P\cdot(\mathcal{M}_\mathrm{enc}(\rho))) =
U\rho U^\dag~,
\end{equation}
and is explicitly referred to by us in~\cite{GHT2005} as the ``encoded
{\em {ersatz}} quantum computer condition," (e$\mathcal{E}$QCC) with the
understanding that the word {\em ersatz} would properly be understood
as meaning {\em inferior} or {\em false}.  Our paper clearly shows
that eq.(8) of~\cite{GHT2005} is not fundamental in any
respect. Indeed, eq.(8) of~\cite{GHT2005} was specifically and
intentionally presented for the sole purpose of illustrating an {\em
{unacceptable}} dynamical condition. This is explicitly pointed out in
several places in our paper, such as in the following sentence ({\em
{cf}} ~the first sentence in the second paragraph above Section 2.3 of
~\cite{GHT2005}): ``However, eq.(8) does {\em not} (emphasis added) in
general provide an acceptable condition to connect the dynamics of a
practical quantum computing device to the constraints implied by the
unitary operator $U$ that defines the abstract quantum computation."
In fact, the various results that were presented by us in~\cite{GHT2005} do
not and indeed {\em cannot} follow from the encoded {\em ersatz} quantum computer
condition. This is because the e$\mathcal{E}$QCC lacks any quantification of implementation 
inaccuracy, {\em{i.e.}},
residual errors not fully removed by error correction.  
Rather, the results of our paper \cite{GHT2005}
follow from a different mathematical expression, the {\em proper} 
quantum computer condition (QCC), in which a
parameter, $\alpha$, 
quantifies the implementation inaccuracy.   
This is discussed further in the second Remark section below. 
\hfill $\Box$
\vspace{.4cm}

\paragraph{\bf Remark:} Claims of equivalence between the UNS of ~\cite{KLPLv1}
and the e$\mathcal{E}$QCC of ~\cite{GHT2005} are irrelevant. None of the 
results of~\cite{GHT2005} could possibly
flow from~\cite{KLPLv1} since any suggested equivalence between the papers 
stems from the misidentification in ~\cite{KLPLv2} of the fundamental expression 
of~\cite{GHT2005}. Nevertheless, even the claimed identification of UNS with the 
e$\mathcal{E}$QCC is wrong as we now show. \hfill $\Box$
\vspace{.4cm}

\paragraph{\bf Response to 2:} The focus of UNS in~\cite{KLPLv1} is the {\em error operators} of 
a quantum communications {\em channel}. This is evident, for example, in the proof of Theorem 6.1 
in~\cite{KLPLv1}, where the Kraus operators associated to the product of $U$ and 
$\mathcal{E}$ are specifically referred to as ``noise 
operators." Note also the many places 
in \cite{KLPLv1} where $\mathcal{E}$ is referred 
to as the ``channel."  
In complete contrast, the e$\mathcal{E}$QCC of \cite{GHT2005} is constructed from operators that 
describe the 
{\em global evolution} of a quantum {\em computer}. 
Any attempt to arbitrarily promote the error operators appearing in \cite{KLPLv1} to 
global evolution operators (note that global evolution includes errors as a special 
case, but not {\em vice versa}) would completely change the original 
meaning and scope of the UNS expression \cite{QE}. (The attempt made in \cite{KLPLv2} to replace 
an error operator with a global evolution operator is described in the next paragraph.)  
This could only be regarded as an attempt after-the-fact to emulate the physical content of the e$\mathcal{E}$QCC.

Apart from the above, the claimed reduction outlined in \cite{KLPLv2} is not mathematically valid. 
According to~\cite{KLPLv2} $P$ acts on the 
computational Hilbert space, $\mathcal H_{\mathrm {comp}}$ ({\em cf} ~the sentence 
following eq.(19) of ~\cite{KLPLv2}: ``\ldots a Hilbert space $\mathcal H_{\mathrm {comp}}$, 
on which $P$ is a quantum operation.").
The authors of \cite{KLPLv2} introduce the composition
$\mathcal{E} = \mathcal{M}_\mathrm{dec} \circ P \circ
(\mathcal{M}_\mathrm{enc} \oplus \opr{id}_\mathcal{K})$ to effect their claimed reduction
({\em cf} ~the last sentence in the second paragraph 
below eq.(19) in~\cite{KLPLv2}). 
However, in this composition the operator $P$ acts 
on the Hilbert space $\mathcal H_{\mathrm {comp}}\oplus \mathcal{K}$,
since $\mathcal{M}_\mathrm{enc}: \mathcal{B}(\mathcal H_{\mathrm {logical}})\rightarrow\mathcal{B}(\mathcal H_{\mathrm {comp}})$ 
({\em cf} the last sentence in the paragraph containing eq.(19) in ~\cite{KLPLv2}). This makes the above 
composition inconsistently defined.

In the converse direction, the authors of~\cite{KLPLv2} give the putative instantiation
$\mathcal{M}_\mathrm{enc}(\sigma^B) = \one^A \otimes \sigma^B$ ({\em cf} the last sentence 
in the last paragraph of Section 4 of~\cite{KLPLv2}).
However, the actual presentation in our paper ~\cite{GHT2005} explicitly refers to 
$\mathcal{M}_\mathrm{enc}$ 
and $\mathcal{M}_\mathrm{dec}$ as ({\em cf} second sentence before eq. (12) in ~\cite{GHT2005}) 
``completely-positive, trace-preserving encoding and decoding maps (with {\em no further 
restrictions of any kind})." (Italics have been added here to the original text of \cite{GHT2005}.) The necessity in the claimed instantiation given in~\cite{KLPLv2} 
to restrict operators  $\mathcal{M}_\mathrm{enc}$ to a particular structure shows that the UNS equation 
is actually a special case of the e$\mathcal{E}$QCC.

It should come as no surprise that the UNS equation is a special case of the e$\mathcal{E}$QCC.  
After all, the e$\mathcal{E}$QCC describes a quantum {\em computer} with perfect error correction, while 
the UNS equation describes a particular type of quantum {\em channel} with perfect error recovery, 
and a quantum 
channel is effectively a quantum computer that is intended to implement the identity 
operation.  

In fact, the reduction of the e$\mathcal{E}$QCC to the UNS equation can be shown explicitly 
by reference to Section 2.3.3 of our paper~\cite{GHT2005}, 
where we demonstrated that 
the correctability criterion used in OQEC ({\em{cf}} \cite{KLPLv1}) is a special case of the 
{\em proper} QCC given 
in eq.(12) of~\cite{GHT2005}.  
We did this by 
first setting our parameter $\alpha$ to zero, which has the effect, through the removal of a norm, of reducing the QCC to the e$\mathcal{E}$QCC. We then showed explicitly how to specialize the operators appearing in the 
e$\mathcal{E}$QCC so as to obtain the correctability criterion of OQEC~\cite{semantic_integrity}.
With this result from \cite{GHT2005}, we now take note of the statement  ({\em cf} ~the first full sentence following  
eq.(18) in~\cite{KLPLv2}) that 
the UNS equation ``is a special case of the OQEC formulation Eq. (13) where 
the recovery $\mathcal{R}$ is unitary" \cite{correctability}.   
Thus we see that the e$\mathcal{E}$QCC is a special case of the QCC, 
the OQEC correctability criterion is a special case of the e$\mathcal{E}$QCC, and
(from the above-cited sentence in \cite{KLPLv2}) the UNS equation is a special case of OQEC 
correctability. Taken together, this establishes the fact that the UNS condition is a special case of the e$\mathcal{E}$QCC.

We re-emphasize that, although Assertion 2 is in fact erroneous, it would be irrelevant 
to any of the results of our paper even if it were true: the mathematical condition from
which the results in our paper {\em actually} derive is the {\em proper} QCC given 
in eq.(12) of~\cite{GHT2005}, as opposed to the clearly labeled encoded {\em ersatz} QCC in 
eq.(8) of ~\cite{GHT2005}.\\
\phantom{~}
\hfill $\Box$
\vspace{.4cm}

\paragraph{\bf Remark:} As a final remark, we note that the generalization of the e$\mathcal{E}$QCC 
(eq.(8) of~\cite{GHT2005}) to the {\em proper} QCC (eq.(12) of~\cite{GHT2005}) involves several nontrivial 
issues, including subtle considerations pertaining to the proper choice of norm. An analysis of some 
of these issues is given in ~\cite{GHTW2005}.
\hfill $\Box$

\bibliography{refs}
\bibliographystyle{hplain}

\end{document}